\documentclass[%
 reprint,
superscriptaddress,
 amsmath,amssymb,
 aps,pre,hyphens,floatfix
]{revtex4-2}

\usepackage[]{amsmath}
\usepackage{amssymb}
\usepackage{color}
\usepackage{enumerate}
\usepackage[usenames,dvipsnames,svgnames,table]{xcolor}
\usepackage{IEEEtrantools}
\usepackage{graphicx}
\usepackage{array}
\usepackage{multirow}
\usepackage{wrapfig}
\usepackage{fullpage}
\usepackage{overpic}
\usepackage{verbatim}
\usepackage{float}
\usepackage{neuralnetwork}
\usepackage[caption=false]{subfig}
\usepackage{multirow}
\usepackage{diagbox}
\usepackage[normalem]{ulem}  

\begin{document}

\preprint{APS/123-QED}

\title{{Identifying Ising and percolation
phase transitions based on KAN method}}

\author{Dian Xu}
\affiliation{Key Laboratory of Quark and Lepton Physics (MOE) and Institute of Particle Physics, Central China Normal University, Wuhan 430079, China}

\author{Shanshan Wang}
\affiliation{Key Laboratory of Quark and Lepton Physics (MOE) and Institute of Particle Physics, Central China Normal University, Wuhan 430079, China}

\author{Wei Li}
\affiliation{Key Laboratory of Quark and Lepton Physics (MOE) and Institute of Particle Physics, Central China Normal University, Wuhan 430079, China}

\author{Weibing Deng}
\affiliation{Key Laboratory of Quark and Lepton Physics (MOE) and Institute of Particle Physics, Central China Normal University, Wuhan 430079, China}

\author{Feng Gao}
\affiliation{Key Laboratory of Quark and Lepton Physics (MOE) and Institute of Particle Physics, Central China Normal University, Wuhan 430079, China}

\author{Jianmin Shen}
\email[]{sjm@mails.ccnu.edu.cn}
\affiliation{Key Laboratory of Quark and Lepton Physics (MOE) and Institute of Particle Physics, Central China Normal University, Wuhan 430079, China}
\affiliation{School of engineering and technology, Baoshan University, Baoshan 678000, China}

\date{\today}

\begin{abstract}

Modern machine learning, grounded in the Universal Approximation Theorem, has achieved significant success in the study of phase transitions in both equilibrium and non-equilibrium systems. However, identifying the critical points of percolation models using raw configurations remains a challenging and intriguing problem. This paper proposes the use of the Kolmogorov-Arnold Network, which is based on the Kolmogorov-Arnold Representation Theorem, to input raw configurations into a learning model. The results demonstrate that the KAN can indeed predict the critical points of percolation models. Further observation reveals that, apart from models associated with the density of occupied points, KAN is also capable of effectively achieving phase classification for models where the sole alteration pertains to the orientation of spins, resulting in an order parameter that manifests as an external magnetic flux, such as the Ising model.

\end{abstract}
\maketitle

\section{Introduction}
\label{intro}


Early investigations into machine learning applications within phase transition models predominantly focused on employing supervised, unsupervised, or semi-supervised methodologies to identify phase transition points in equilibrium systems, non-equilibrium systems, quantum phase transitions, and topological phase transitions\cite{carrasquilla2017machine,wang2016discovering,shen2021supervised,ohtsuki2016deep,deng2017machine,rodriguez2019identifying,ahmed2025supervised,ho2023self}. In recent years, machine learning techniques have increasingly been applied to more intricate phase transition problems, encompassing the exploration of system microstructures through generative models such as variational autoencoders(VAEs)\cite{kingma2019introduction,pu2016variational} and generative adversarial networks(GANs)\cite{goodfellow2020generative}. Additionally, research based on graph neural networks(GNNs)\cite{wu2020comprehensive,ma2023jet} and active learning\cite{ziatdinov2022bayesian,lookman2019active,demir2011detection} is emerging, aimed at addressing physical systems with complex interactions, thereby advancing the deeper integration of machine learning in the realm of phase transitions.

As foundational frameworks for phase transition research, the Ising and percolation models are indispensable. Employed as recent investigative methodologies, machine learning boasts a substantial corpus of research achievements pertaining to these two models.

Wenjian Hu et al. applied principal component analysis (PCA)\cite{abdi2010principal} and autoencoder methods to investigate the phase behavior and phase transitions of several classical spin models, including the square and triangular lattice Ising models, and the two-dimensional XY model. They found that the principal components quantified by PCA not only explore different phases and symmetry breaking but also distinguish types of phase transitions and locate critical points. The study also demonstrated that autoencoders can be trained to identify critical points of phase transitions.\cite{wetzel2017unsupervised} Sebastian J. Wetzel examined unsupervised machine learning techniques, specifically PCA and variational autoencoders based on artificial neural networks, by learning the most descriptive feature configurations of the two-dimensional Ising model and the three-dimensional XY model. It was discovered that the latent parameters correspond to known order parameters.\cite{hu2017discovering}

The order parameter in percolation models is not the density of active sites; it also includes the probability that lattice sites (or bonds) belong to the percolating cluster. As a result, using unsupervised learning to identify the critical point in percolation models has been a persistent challenge. Zhang Wanzhou applied an Ising mapping approach to map the original configurations of the percolation model, subsequently using machine learning to identify the system’s critical point.\cite{zhang2019machine} Shu Cheng and colleagues used various neural networks to study configurations with noise.\cite{cheng2021machine} 

Jianmin Shen’s unsupervised learning results on the 1+1 dimensional directed percolation (DP) model indicate that both the first principal component from PCA and a single latent neuron from an autoencoder effectively represent the DP model’s order parameter, namely, the particle density. Interestingly, when the lattice configurations are randomized, the unsupervised learning results show no difference from those obtained with the original configurations.\cite{jianmin2023machine}

The mathematical foundation of MLPs (Multilayer Perceptrons)\cite{tolstikhin2021mlp,pinkus1999approximation} is the Universal Approximation Theorem, which posits that a feedforward neural network equipped with a sufficient number of neurons can approximate any complex continuous function, irrespective of the number of input variables. The Universal Approximation Theorem underscores the necessity of a sufficiently wide (with n being sufficiently large) network to attain a low error between the output and the target function. Theoretically, an infinite level of precision would require an infinitely wide network.

Various methodologies exist for achieving function approximation, each distinguished by distinct implementation approaches and theoretical underpinnings. Kolmogorov-Arnold Representation Theorem (KA Representation Theorem)\cite{schmidt2021kolmogorov} posits that if $f$ is a continuous multivariate function, it can be expressed as a sum of finitely many univariate functions. 



Inspired by this, in this paper, we employ the KAN\cite{liu2024kan} method to identify phase transitions in the Ising and percolation models and successfully locate the critical points of these models. Subsequently, we utilize finite-size scaling to extrapolate the critical points to the thermodynamic limit. Finally, we determine the critical exponents of the models through data collapse analysis. The effectiveness of the KAN method is substantiated by comparing our findings with results from numerical simulations and theoretical calculations.


The primary structure of this paper is delineated as follows. Section~\ref{Models} elaborates on the Ising model and percolation model of interest. Section~\ref{method} provides a detailed exposition of the KAN method. Section~\ref{usml_result} delineates the research findings, presenting the results analysis of both models under KAN. Conclusively, Section~\ref{Conclusion} furnishes a comprehensive summary of the present study.

\section{Models}\label{Models}
\subsection{percolation model}\label{percolation model}

Research on percolation models has a long history, beginning with studies of how fluids diffuse through the pores of coal.\cite{broadbent1957percolation} Modern percolation theory has evolved to focus on changes in network behavior as nodes or edges are added.\cite{kirkpatrick1973percolation} The work in \cite{shante1971introduction} represents one of the earliest systematic discussions of the physical and geometric properties of percolation models.

\begin{figure*}[htbp]
\begin{tabular}{cc}
    \includegraphics[width=0.35\textwidth]{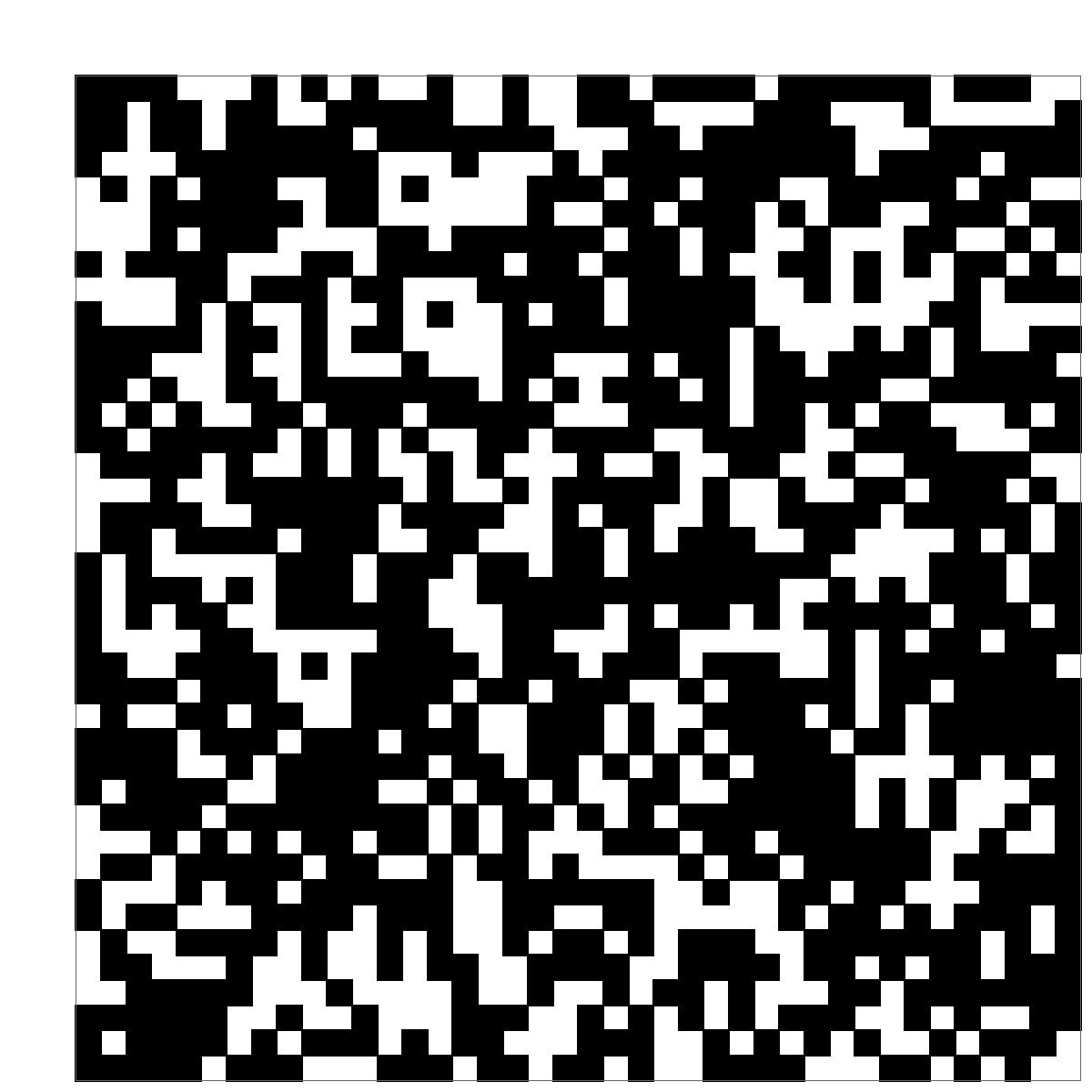} &
    $\qquad$\includegraphics[width=0.35\textwidth]{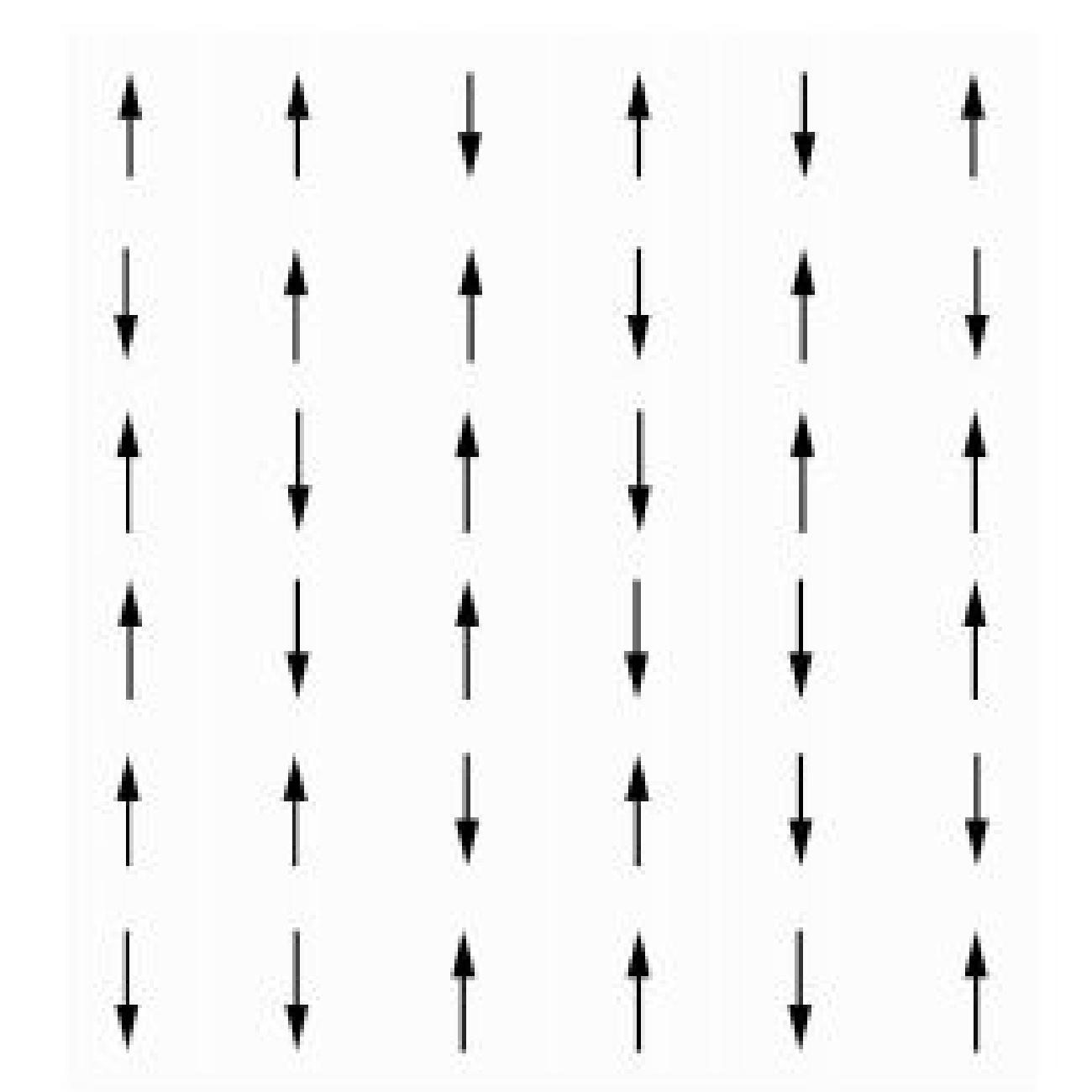} \\
    (a) & $\qquad$ (b)
\end{tabular}
\caption{In this article, we examine two distinct configurations. \textbf{a} is the raw configuration of site percolation with occupation probability = 0.8. \textbf{b} is the raw configuration of ising model.}
\label{percolation&ising_configuration}
\end{figure*}

The two-dimensional percolation model represents a continuous phase transition,\cite{riordan2011explosive} with the order parameter given by
\begin{equation}
   P_{\infty} (p) \propto (p - p_{c})^{\beta}  \qquad for \quad p \rightarrow p_{c} ^{+}
\end{equation}
where $p$ is the occupation probability, $p_{c}$ is the critical probability,commonly referred to as the percolation threshold. $\beta$ is the critical exponent of the order parameter, and $P_{\infty} (p)$ denotes the probability that a given site belongs to the percolating cluster namely percolation probability. Typically, numerical simulation methods are employed to compute $P_{\infty} (p)$ to locate the critical point.

In the context of a square lattice, the probability of encountering a percolating cluster within the system remains extremely low as long as the probability remains below the critical probability $p_c$. Conversely, when the probability surpasses the critical probability $p_c$, the likelihood of encountering a percolating cluster rapidly approaches unity as the number of occupied sites increases.



To ascertain the critical point of the two-dimensional site percolation model, a statistical description of finite cluster sizes can be derived based on cluster density. The precise expression for cluster density is as follows.
\begin{equation}
    n(s,p)=\sum\limits_{t=1}^{\infty}g(s,t)(1-p)^{t}p^{s}
\end{equation}
Here, $g(s,t)$ represents the number of distinct clusters of size $s$ and perimeter $t$. Unfortunately, a unique relationship between size $s$ and perimeter $t$ does not exist. For smaller clusters, $g(s,t)$ can be manually enumerated, but as cluster size increases, the expression for $n(s,p)$ becomes increasingly complex. Therefore, we can only resort to numerical simulations.

Within the framework of the site percolation model, an initial step involves constructing a lattice composed by $N \times N$ grid points. Subsequently, each grid point is randomly occupied with a specified probability $p$. Adjacent occupied grid points are considered to be part of the same cluster. The occurrence of a percolation phase transition is identified when a cluster spans the entire lattice. To obtain statistically robust data, multiple repetitions of this simulation are essential.


For cluster size and cluster density, experimental results reveal a highly linear relationship on a log-log scale at a probability $p_{c} = 0.592746$\cite{christensen2005complexity}.
\begin{equation}
    n(s,p)\propto s^{-\tau}
\end{equation}

In the two-dimensional site percolation model, the critical exponent $\tau$ is determined to be 187/91.


\subsection{Ising model}\label{ising model}
The Ising model represents a highly simplified model in relation to physical reality, employed to elucidate the ferromagnetic properties of materials\cite{brush1967history,newell1953theory}. Specifically, a two-dimensional square lattice Ising model stands as the simplest known physical system that exhibits a phase transition\cite{mccoy1973two}.

In this model, a parameter $\sigma_{i}$ is incorporated to characterize the magnetic moment of individual atoms, with values restricted to +1 or -1, signifying spin-up or spin-down orientations, respectively. For two adjacent lattice sites $i$, $j$ $\in$ $\Lambda$, an interaction parameter $J_{ij}$ is introduced. Consequently, the Hamiltonian of the entire system can be expressed as:
\begin{equation}
    H(\sigma)=-\sum \limits_{<i,j>}{J}_{ij}\sigma _{i}\sigma _{j}-\mu \sum \limits_{j}h_{j}\sigma _{j}
\end{equation}

Here, $<i,j>$ signifies that the lattice points $i$ and $j$ are adjacent. Consequently, the first term of the Hamiltonian represents the sum over all pairs of neighboring lattice points (with each pair counted only once), corresponding to the energy of interactions between all spins. The second term denotes the energy of interaction between the magnetic field and the spins. The parameter $\mu$ represents the magnitude of the magnetic moment at each lattice point.

The configurational probability $P(\sigma)$ of the system represents the likelihood of a specific spin configuration $\sigma$ occurring under thermal equilibrium, adhering to the Boltzmann distribution:
\begin{equation}
    P_{\beta}(\sigma)=\frac{e^{-\beta H(\sigma)}}{Z_{\beta}}
\end{equation}

Here,$\beta=(k_{B}T)^{-1}$, and
\begin{equation}
    Z_{\beta}=e^{-\beta H(\sigma)}
\end{equation}

The term serves as the normalization constant of the probability distribution, a concept also known as the partition function in statistical mechanics. For a physical quantity \( f(\sigma) \) that is a function of the spin configuration \(\sigma\), its expected value can be expressed as:
\begin{equation}
    <f>_{\beta}=\sum\limits_{\sigma}f(\sigma)P_{\beta}(\sigma)
\end{equation}

A common simplification involves assuming no external magnetic field is applied to the Ising model and that all interactions between neighboring lattice sites are equal, i.e., $h_{j}$ = 0 and $J_{ij}$ = $J$ for all $i$, $j$ $\in$ $\Lambda$. Under this simplification, the Hamiltonian can be expressed as:
\begin{equation}
    H(\sigma)=-J\sum\limits_{<i,j>}\sigma_{i}\sigma_{j}
\end{equation}

At this juncture, the Ising model exhibits symmetry under the operation of inverting all spins\cite{cipra1987introduction}.

Consider a two-dimensional square lattice Ising model with interaction energies along the horizontal and vertical directions given by \( J1 \) and \( J2 \), respectively. Lars Onsager derived the analytical expression for the free energy in the absence of an external magnetic field\cite{onsager1944crystal}, i.e., when \( h = 0 \).
\begin{equation}
\begin{split}
        -\beta f=ln2+\frac{1}{8\pi^{2}}\int_{0}^{2\pi}d\theta_{1}\int_{0}^{2\pi}d\theta_{2}  \\
        ln[cosh(2\beta J_{1})cosh(2\beta J_{2})-  \\  sinh(2\beta J_{1})cos(\theta_{1})-sinh(2\beta J_{2})cos(\theta_{2})]
\end{split}
\end{equation}

Various thermodynamic functions can be obtained from the partial derivatives of the Gibbs free energy.Specifically, the two-dimensional Ising model has a critical point at the critical temperature \( T_c \), which satisfies the following equation:
\begin{equation}
    sinh(\frac{2J_{1}}{kT_{c}})sinh(\frac{2J_{2}}{kT_{c}})=1
\end{equation}

If we set $J_{1}=J_{2}$ and $k=1$, then we get critical temperature
\begin{equation}
    T_{c}=\frac{2J}{kln(1+\sqrt{2})}=2.27
\end{equation}

\section{KAN}\label{method}

\begin{figure*}[htbp]
\centering
\includegraphics[width=0.8
\textwidth]{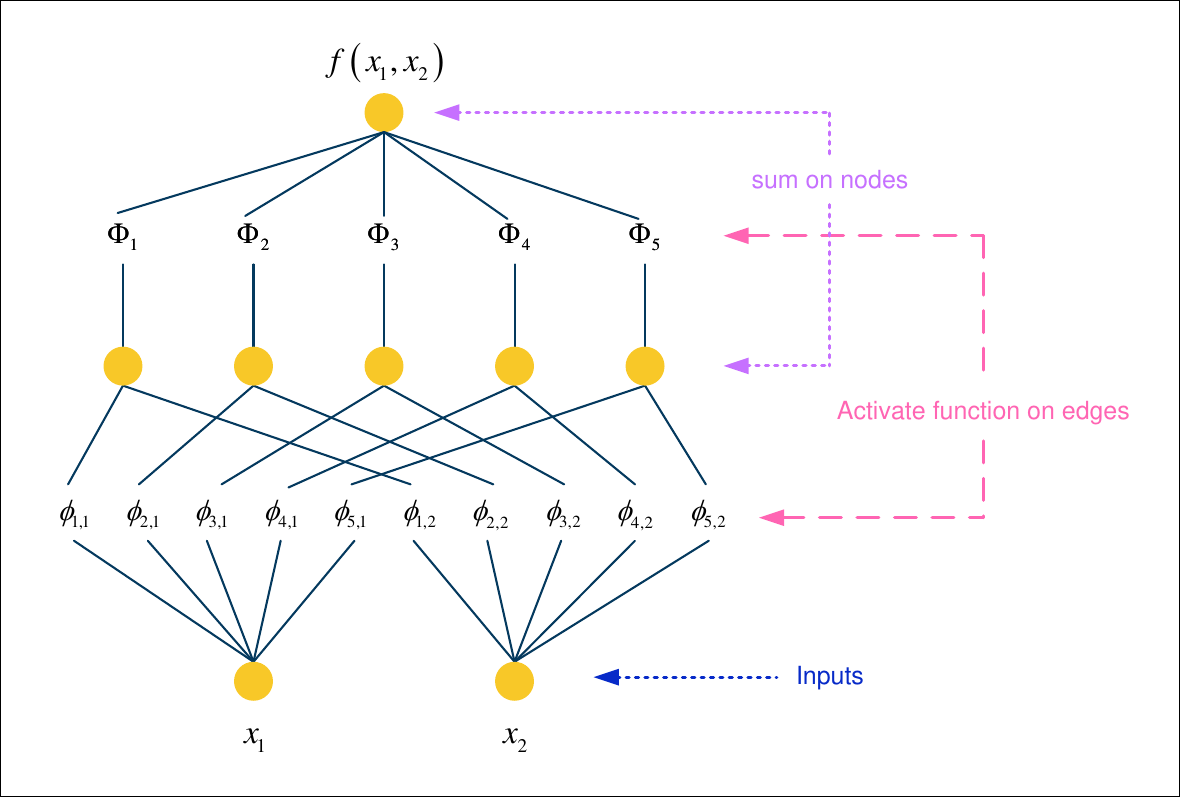}
\caption{Neural network schematic structure of KAN.}
\label{fig:kan}
\end{figure*}




The Kolmogorov-Arnold Network (KAN) as shown in Fig\ref{fig:kan} is inspired by the Kolmogorov-Arnold Representation Theorem (KA Representation Theorem), as demonstrated by the equation, asserts that an approximation with infinite precision can be achieved using a network of finite size, albeit such an approximation may necessitate the use of non-smooth activation functions.
\begin{equation}
    f(x_1, x_2, ... x_n)=\sum\limits_{j=1}^{2n+1}\psi_q (\sum\limits_{i=1}^{n}\psi_{qp} (x_i))
\end{equation}

Based on the aforementioned formula, it is evident that in order to transform it into a neural network, the first layer would consist of $n$ neurons, the second layer would contain $2n+1$ neurons, and the third layer would have $1$ neuron. This configuration represents the original kan network, which may involve some pathological activation functions designed to approximate the final representation function. However, our observation reveals that if a highly pathological activation function exists, it can be decomposed using some smooth activation functions. This insight allows for the deepening of the original kan network, enabling the original KA representation theorem to be rewritten in the following form.
\begin{equation}
\begin{split}
    f(x)=\sum\limits_{i_{L-1}=1}^{n_{L-1}}\psi_{L-1,i_{L},i_{L-1}}(\sum\limits_{i_{L-2}=1}^{n_{L-1}} ...(\sum\limits_{i_{2}=1}^{n_{2}}\psi_{2,i_{3},i_{2}} \\ (\sum\limits_{i_{1}=1}^{n_{1}}\psi_{1,i_{2},i_{1}}(\sum\limits_{i_{0}=1}^{n_{0}}\psi_{0,i_{1},i_{0}}(x_{i_{0}}))))...)
\end{split}
\end{equation}

In the Kan model, nodes exclusively perform summation operations, while the transformation from $x$ to $\phi(x)$ is executed along the edges. Specifically, the activation function is applied on the edges.
\begin{equation}
    \psi(x)=b(x)+spline(x)
\end{equation}
While $b(x)$ is silu function.
\begin{equation}
    b(x)=\frac{x}{1+e^{-x}}
\end{equation}

The spline function, denoted as $spline(x)$, can be expressed as 
\begin{equation}
    spline(x)=\sum\limits_{i=1}^{n+1} c_{i}B_{i}(u)
\end{equation}

The basis functions of the B-spline function can be defined through the De Boor-Cox recursion, formulated as follows:
\begin{equation}
    B_{i,0}(u)=\left\{
\begin{aligned}
1 && for && u_{i}\leq u \leq u_{i+1} \\
0  && others
\end{aligned}
\right.
\end{equation}
\begin{equation}
    \begin{split}
    B_{i,k}=\frac{u-u_{i}}{u_{i+k}-u_{i}}B_{i,k-1}(u)+ \\
    \frac{u_{i+k+1}-u}{u_{i+k+1}-u_{i+1}}B_{i+1,k-1}(u)  \cr  for \quad  u_{k}\leq u \leq u_{n+1}
    \end{split}
\end{equation}

We have identified three unknown parameters in this context: $m+1$ signifies the number of basis functions employed, $n+1$ denotes the number of control points selected, and $k$ represents the order of the spline basis. These parameters are subject to a rather weak constraint, as delineated below.
\begin{equation}
    m=n+k+1
\end{equation}

The utilization of B-splines for approximating the target function offers substantial flexibility. To ascertain the unique form of the coefficients $c_i$, it is imperative to impose three constraints on the control points. 

1. Clamped B-spline Curves Condition: This involves repeating the first and last knots $k+1$ times, thereby ensuring that the generated B-spline curves pass through the initial and terminal data points.

2. Uniform Knot Distribution Condition: This stipulates that, apart from the first and last knots, the intervals between control points are equidistant.

3. Non-knot Boundary Condition: This condition requires that the third derivative at the first knot equals that at the second knot, and similarly, the third derivative at the nth knot equals that at the (n+1)th knot.

The ultimate goal is to minimize the $L_{total}$. 
\begin{equation}
L_{total}=L_{pred}+\lambda(\mu_{1}\sum\limits_{l=1}^{L}|\Phi_{l}|_{1}+\mu_{2}\sum\limits_{l=1}^{L}S(\Phi_{l}))
\end{equation}
where $\mu_{1}$,$\mu_{2}$ are relative magnitudes usually set to $\mu_{1}$=$\mu_{2}$=1, and $\lambda$ controls overall regularization magnitude.

The first term on the right-hand side of the above equation represents \( L_{pred} \). It is cross-entropy loss function between the predicted output $T$ and the target input $R$, which will lead to the network’s output. It is defined by the following formula:
\begin{equation}
    H(R,T) = \sum R(x_i) log \frac{1}{T(x_i)}
\end{equation}

The second part is 'active function'-entropy, which will minimal network structure. First We define the $L1$ norm of an activation function $\psi$ to be its average magnitude over its $N_{p}$ inputs,
\begin{equation}
    |\psi|_{1}=\frac{1}{N_{p}}\sum\limits_{s=1}^{N_{p}}|\psi(x^{(s)})|
\end{equation}

Then for a KAN layer $\Phi$ with $n_{in}$ inputs and $n_{out}$ outputs, we define the L1 norm of $\psi$ to be the
sum of L1 norms of all activation functions,
\begin{equation}
|\Phi|_{1}=\sum\limits_{i=1}^{n_{in}}\sum\limits_{j=1}^{n_{out}}|\psi_{i,j}|_{1}
\end{equation}

In addition, we define the entropy of $\Phi$ to be
\begin{equation}
S(\Phi)=-\sum\limits_{i=1}^{n_{in}}\sum\limits_{j=1}^{n_{out}}\frac{|\psi_{i,j}|_{1}}{|\Phi|_{1}}log(\frac{|\psi_{i,j}|_{1}}{|\Phi|_{1}})
\end{equation}

Through the utilization of backpropagation and gradient descent algorithms, we are capable of updating the parameters within the activation function, specifically the amplitudes \( c_{i} \) of the basis functions in the spline function. This process culminates in optimizing the learning outcome. By doing so, we ensure that the cross-entropy between the input and output is minimized, while simultaneously maintaining a simplified network structure through the employment of a minimal number of activation functions. Specifically, the update rule is:
\begin{equation}
    c_{i} \leftarrow c_{i}-\eta\frac{\partial L}{\partial c_{i}}
\end{equation}
where $\eta$is the learning rate, $\frac{\partial L}{\partial c_{i}}$ is the gradients of the loss function $L$ with respect to the amplitudes \( c_{i} \).

\section{The KAN Learning Results}
\label{usml_result}

In the preceding section, we delineated the models under investigation and the machine learning methodologies employed. This segment will meticulously dissect the outcomes of the analysis and the procedural steps leading to these results. Initially, we generated the raw data for the models in question through the Monte Carlo method. For the two-dimensional Ising model, the Metropolis algorithm was utilized to generate the data, with the temperature set at 101 distinct values ranging from 0.02 to 4.02. A square lattice was employed, coupled with periodic boundary conditions. At each temperature point, a sample size of 1,000 was maintained, resulting in the data being stored as 101 three-dimensional arrays of dimensions [1000, L, L]. Subsequently, the data was flattened into a two-dimensional array of shape [1000, L*L], which was then fed into the KAN network.

Similarly, for the percolation model, we set the occupation probability of sites from 0 to 1, encompassing 101 discrete values.

In the KAN network, the number of nodes in the input layer is set to $l$*$l$, while the hidden layer, which is the second layer of the network, is configured with 5 nodes (the rationale behind this specific number has not been theoretically substantiated prior to the completion of this manuscript). To achieve the ultimate goal of phase classification, the output layer is designed with 2 nodes.

To investigate the impact of system size on critical point measurements, both the Ising and percolation models were configured with system sizes of 10, 20, 30, and 40.

Having elucidated the methodology of employing the kan network for model analysis, we present a concise overview of the training accuracy achieved by the model under varying system sizes and training steps, as detailed in Table \ref{acc}. It is readily apparent from the table that the training accuracy of the kan model progressively improves with the escalation of system size and the augmentation of training steps. Consequently, in subsequent experiments, we standardize the training steps to 200 across all system sizes to ensure the reliability of experimental outcomes.

\begin{table*}[tbh!]
	\centering
	\begin{tabular}{|c|c|c|c|c|c|c|}
        \hline
       \diagbox[width=10em]
       {L}{Accuracy}{Steps}     
       &50 for Ising &100 &200 &50 for percolation &100 &200    \\
        \hline
   10   &0.752  &0.892  &0.995   &0.693   &0.894  &0.992  \\
        \hline
   20   &0.844  &0.993  &0.994   &0.912   &0.978  &0.991   \\
        \hline
   30   &0.942  &0.957  &0.998   &0.924   &0.956  &0.998   \\
        \hline
   40   &0.992  &0.994  &0.997   &0.995   &0.997  &0.999   \\
        \hline
   \end{tabular}
\caption{
This table presents the accuracy rates of the Ising and percolation models across varying dimensions and training steps. The rightmost column signifies the size of the models, while columns 2 through 4 depict the evolution of accuracy in the Ising model as training steps increase. Columns 5 to 7 reflect the corresponding accuracy changes in the percolation model.}
\label{acc}
\end{table*}

\subsection{Detect crital point}\label{singmoid fit}

\begin{figure*}[t]
\begin{tabular}{cccc}
    \includegraphics[width=0.45\textwidth]{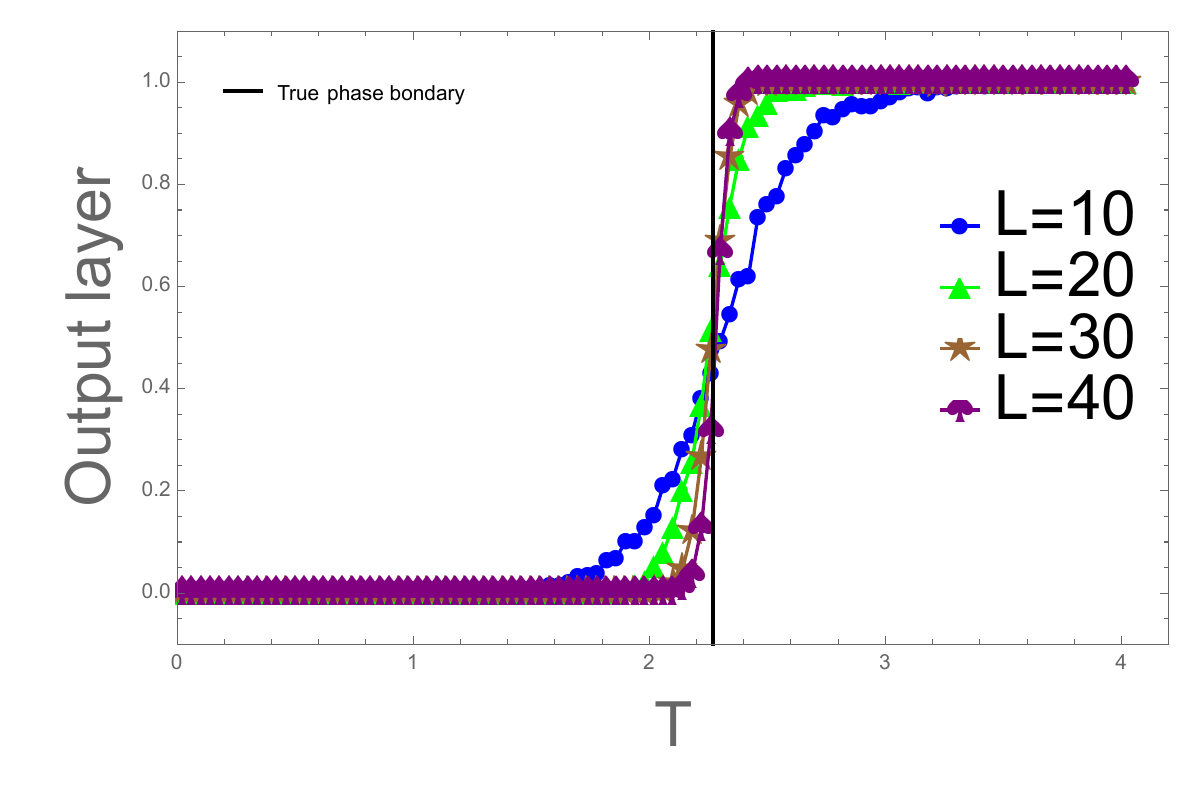}&
    \includegraphics[width=0.45\textwidth]{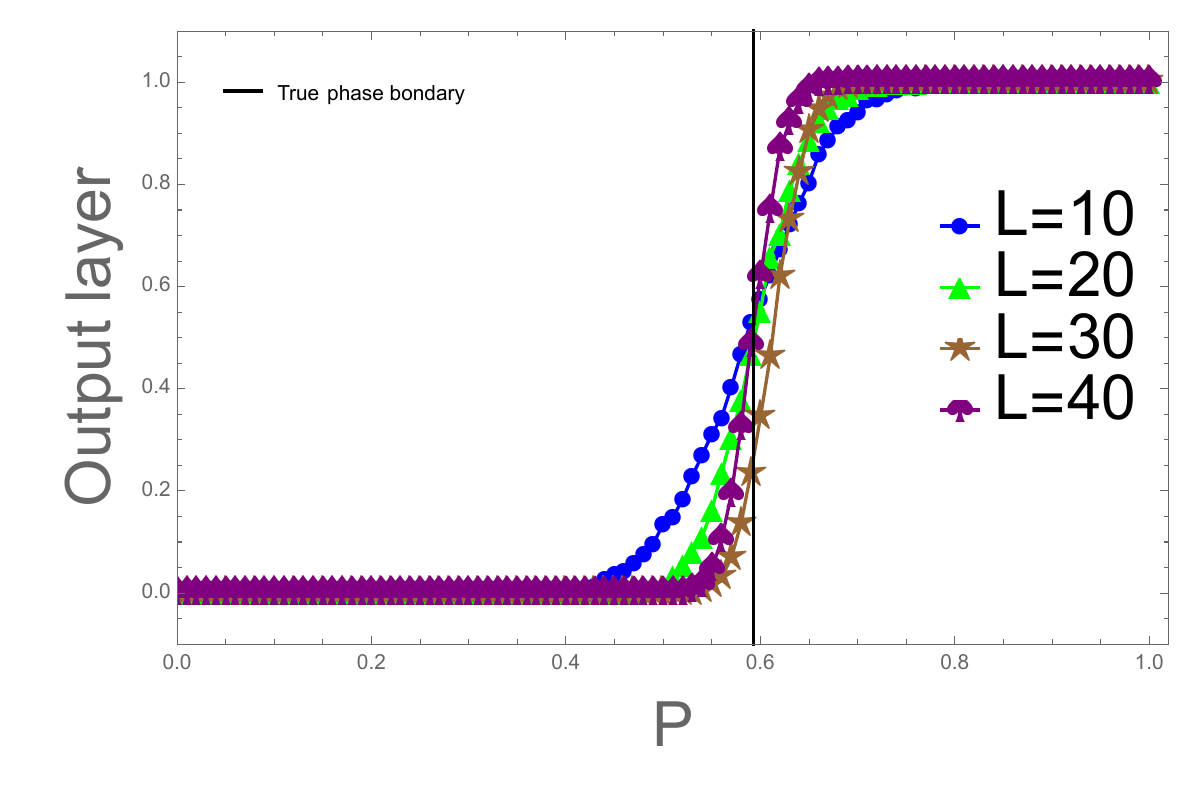}&  \\
    (a) &  (b)    \\
    \includegraphics[width=0.45\textwidth]{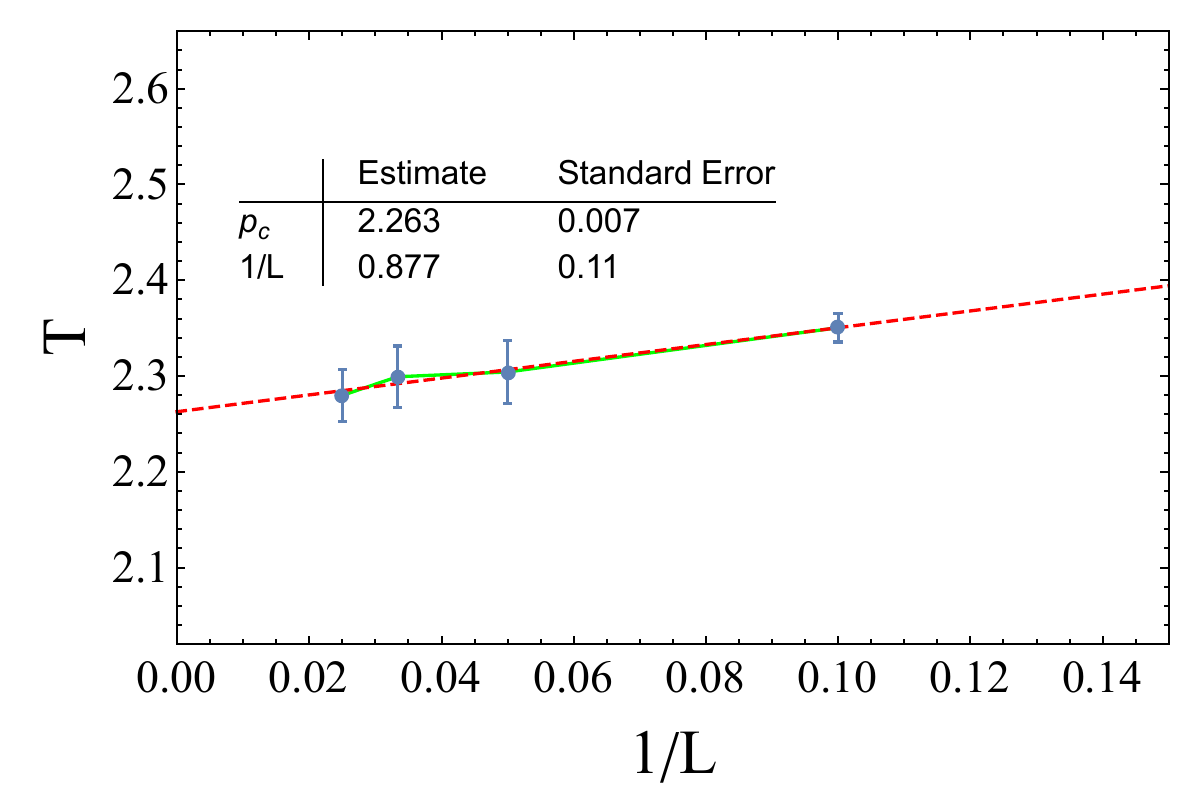} &
    \includegraphics[width=0.45\textwidth]{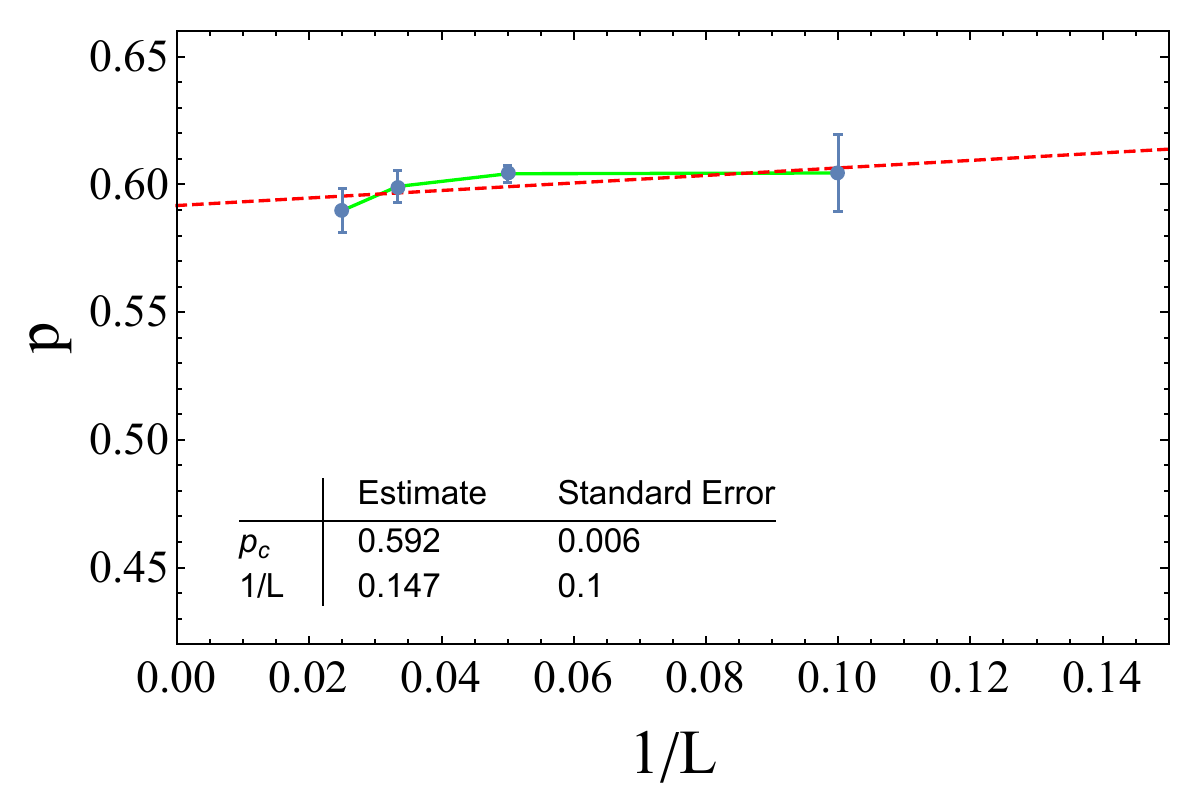}&  \\
    (c) &  (d) 
\end{tabular}
\caption{
(a)Presents the training outcomes of the Ising model under various dimensions, denoted as kan. (b) Depicts the results of site percolation. Both scenarios encompass sizes of 10, 20, 30, and 40, with a sample size of 1000 for each. (c) Illustrates the extrapolated results of the Ising model to the thermodynamic limit. (d) Represents the outcomes of percolation in the same thermodynamic limit.}
\label{jieguo}
\end{figure*}

For the Ising model and the Percolation model, we employed the learning outcomes of the KAN network, as depicted in Figure \ref{jieguo}(a)(b). It is evident from the figure that the KAN network achieves an excellent classification of the configurations with respect to temperature or probability variations. As the temperature or probability increases, the steady-state configurations of the models transition from one state to another, exhibiting a step-function-like pattern. This transition is marked by a distinct jump behavior at the critical point, which aligns with our expectations. Subsequently, we applied the sigmoid function for fitting, with the functional form as expressed in the following equation, thereby quantitatively determining the critical point.
\begin{equation}
    f(x)=\frac{1}{1+e^{-k(x-x_{0})}}
    \label{sigmoid function}
\end{equation}

Subsequent to this, we conducted multiple experiments, employing the method of averaging measured values to minimize the random errors inherent in single measurements. Additionally, we augmented the sample size to mitigate the impact of systematic thermal fluctuations on the experimental outcomes, thereby reducing associated errors.

Finally, taking into account the influence of system size on critical point measurements, we employed finite-size scaling techniques. By extrapolating the system size, we obtained the theoretical critical point value of the model under thermodynamic limit conditions, as illustrated in Figure \ref{jieguo}(c)(d). The abscissa represents the reciprocal of the model size, where $1/L$=0 signifies an infinitely large model size, and the ordinate corresponds to the critical point value of the model. It is evident from the figure that the critical point values obtained through this fitting method are remarkably close to the theoretical critical point values\cite{onsager1944crystal,christensen2005complexity}, which substantiates the effectiveness of the Kan method.

\subsection{Data collapses for critical exponent}

In the preceding section, we measured the critical point of the model; however, for a phase transition system, it is equally crucial to ascertain its universality class, which is typically characterized by critical exponents. Employing the method of data collapse, we can effectively determine the numerical values of these critical exponents.


Prior to determining the critical exponents, it is imperative to delineate the system's correlation function, which elucidates the correlation between two points within the model. Its definition is as follows.
\begin{equation}
    C_{\perp}(r) = C_{\perp}(t,r) = \langle s_{i}(t)s_{i+r}(t) \rangle.
\label{corr_function}
\end{equation}

\begin{figure*}[!htb]
\begin{tabular}{ccc}   
    \includegraphics[width=0.45\textwidth]{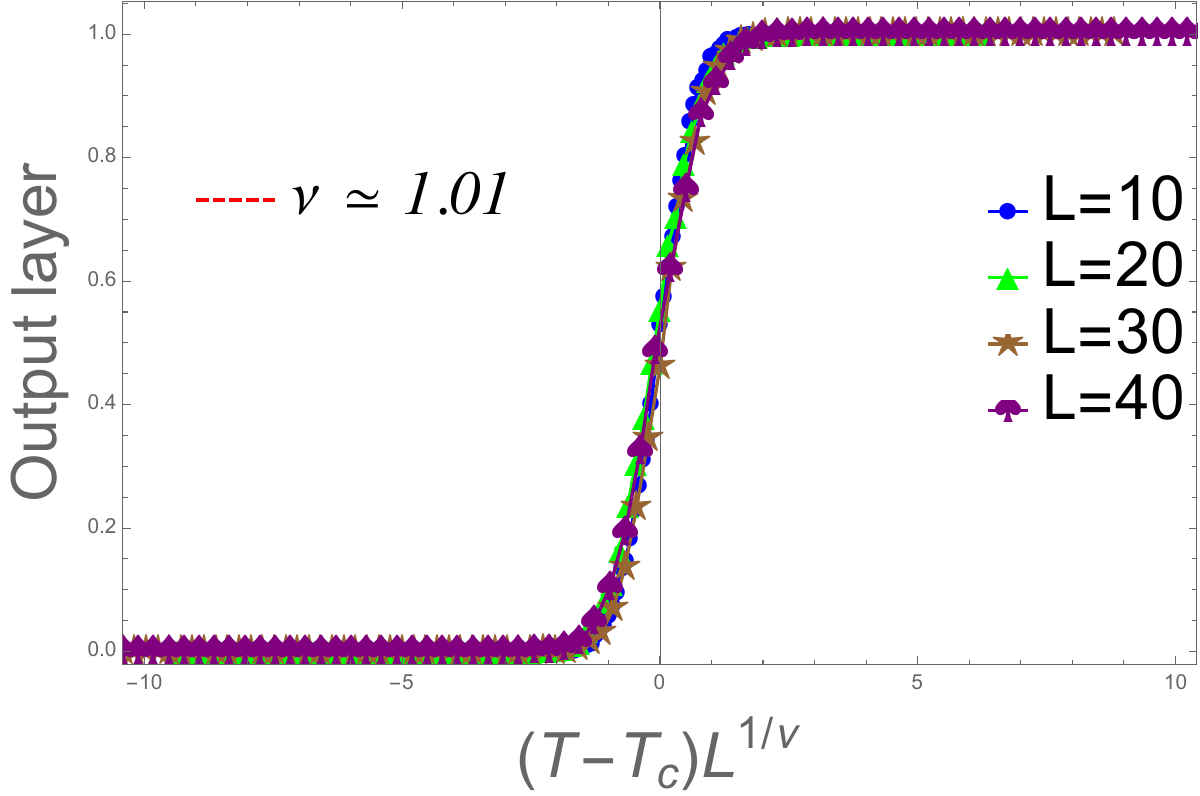} &
    \includegraphics[width=0.45\textwidth]{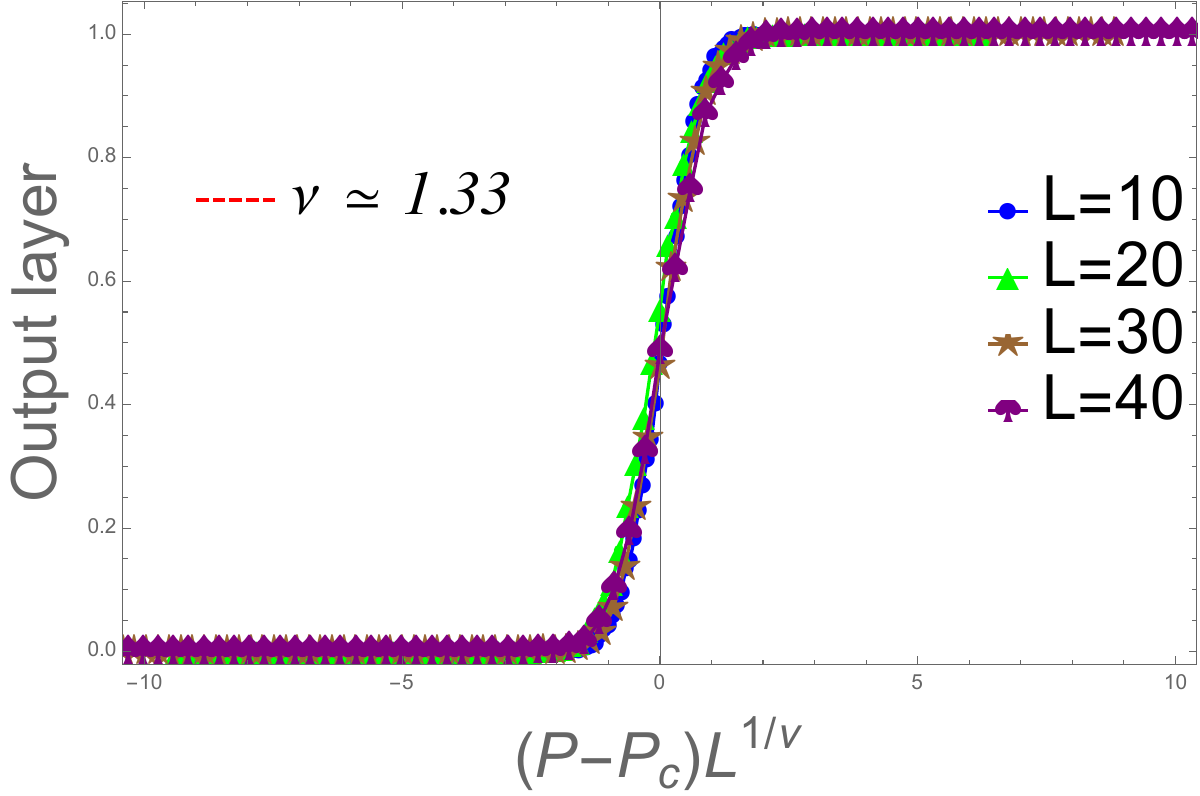}\\
     (a) &  (b)
\end{tabular} 
\caption{
The experimental results of data collapse. Panel \textbf{a} shows the results of Ising, and panel \textbf{b} shows the results of Percolation.The collapse of the average output layer as a function of $(T- T_{c}) L^{1/ \nu_{\perp}}$ and $(P- P_{c}) L^{1/ \nu_{\perp}}$.}
\label{datacop}
\end{figure*}

Through mathematical transformation and the redefinition of parameters, the correlation function can be expressed in the following form.
\begin{equation}
C(r)\sim exp (- \frac{r}{\xi}),
\label{corr_func}
\end{equation}

Here, $\xi$ denotes the correlation length, which serves to delineate the extent of correlation present among magnetic moments or particle fluctuations. Specifically, when the distance $r$ is less than $\xi$, the correlation of fluctuation amplitudes is notably significant. Conversely, when $r$ exceeds $\xi$, the correlation function rapidly decays to zero. Experimental findings indicate that in the vicinity of the critical point, the relationship between the correlation length and temperature or probability can be expressed as:
\begin{align}
    \xi \sim \left| T- T_{c} \right|^{-\nu_1}  \\
    \xi \sim \left| P- P_{c} \right|^{-\nu_2}
\label{nu}
\end{align}

Here, \(\nu\) represents a critical exponent, which may be identical across various phase transition models, thereby enabling us to categorize universality classes based on distinct critical exponents. Concurrently, it is established that a proportional relationship exists between the correlation length and the system size, allowing the aforementioned formula to be reformulated as follows.
\begin{align}
    \xi \sim L \sim \left| T- T_{c} \right|^{-\nu_{1}} \rightarrow\left| T- T_{c} \right| \sim L^{-1/\nu_1}  \\
    \xi \sim L \sim \left| P- P_{c} \right|^{-\nu_{2}} \rightarrow\left| P- P_{c} \right| \sim L^{-1/\nu_2}
\label{per}
\end{align}

Thus, we establish a connection between the critical exponents and the observable physical quantities, ultimately determining the numerical values of the critical exponents through the method of data collapse.

The results of critical exponents obtained through data collapse are illustrated in Figure \ref{datacop}. Panel $a$ presents the outcomes for the Ising model, where the data collapse method has successfully achieved a satisfactory fit to the four system size-similar curves depicted in Figure 8(c), yielding a critical exponent of $\nu$ = 1.01. This numerical solution aligns well with the results reported in the literature\cite{boulatov1987ising}. Panel $b$ showcases the results for the percolation model, demonstrating a similarly effective curve fitting, resulting in a critical exponent of $\nu$ = 1.33, which is in close proximity to the values documented in the literature\cite{smirnov2001critical}.

\subsection{Conclusion}\label{Conclusion}
In this paper, we utilized KAN to conduct an in-depth investigation of the critical behavior of two representative phase transition models: the site percolation model in geometric phase transitions and the two-dimensional Ising model as a classical example of equilibrium phase transitions. Notably, previous attempts to locate the critical point of the percolation model through unsupervised learning on the original configurations were unsuccessful. This paper proposes the use of KAN to analyze the original configurations of percolation, successfully identifying the critical point of the phase transition. Here, KAN is employed as a non-global approximation method, capable of predicting key information about critical behavior by labeling part of the original configurations as training inputs. This approach extends the application scope of machine learning (ML) techniques in the study of critical behavior.

By evaluating the output labels from KAN, we predicted the critical points of the system under different lattice sizes. Subsequently, we obtained the theoretical values for an infinite system using a fitting method based on finite-size scaling. Furthermore, the phase score curves can be fitted to an S-like function, which allows us to compute the critical exponent $\nu$ for site percolation and the two-dimensional Ising model in a phase transition system through data collapse.

\section{Discussion}
This paper analyzes the raw configurations of the Ising and percolation models using the KAN. The operation principle of the network is based on the Kolmogorov's representation theorem. However, whether the Universal Approximation Theorem of the MLP is applicable to the raw configurations remains open to debate. 

The literature \cite{alexandrou2019unsupervised} employed autoencoders (AE) to experimentally analyze the original configurations of the Ising model, yielding results similar to the sigmoid function, with the critical point determination consistent with theoretical values. Similarly, Wang L used the principal component analysis (PCA) method in a 2016 paper to determine the critical point of the original configurations of the Ising model\cite{wang2016discovering}.

In our previous work, we discussed how the critical point of the percolation model could be identified through its largest cluster\cite{xu2025identifying}. However, it is intriguing that when using the raw configurations to locate the numerical value of the critical point, both PCA and AE only yield a linear relationship between the density of active sites and the occupation probability, which is of no help in determining the numerical value of the critical point.

\section{Acknowledgements}

This work was supported in part by Research Fund of Baoshan University(BYKY202305), Yunnan Fundamental Research Projects (Grant 202401AU070035), the Fundamental Research Funds for the Central Universities, China (Grant No. CCNU19QN029), the National Natural Science Foundation of China (Grant No. 61873104), the 111 Project, with Grant No. BP0820038.


The detailed algorithms of how to generate data and use machine learning are shown in the GitHub link {https://github.com/freeupcoming/kan-Ising-percolation}.

\bibliographystyle{apsrev4-2}
\bibliography{sitepercolation}

\end{document}